\documentclass[aps,pra,twocolumn,amssymb,amsfonts,floatfix]{revtex4}
\usepackage{bm}
\usepackage{dcolumn}
\usepackage{bm}
\usepackage{mathbbol}
\usepackage{graphicx}
\usepackage{epstopdf} 
\usepackage{amsmath,amsthm}
\usepackage{times}
\usepackage{color}
\usepackage{lipsum}

\begin{document}
\title{Effects of excited state quantum phase transitions on system dynamics}
\author{Francisco P\'erez-Bernal}
\affiliation{Grupo de investigaci\'on en F\'{\i}sica Molecular, At\'omica y nuclear (GIFMAN-UHU), Unidad Asociada al CSIC. Departamento de Biolog\'{\i}a, F\'{\i}sica y Matem\'aticas, Universidad de Huelva, 21071 Huelva, SPAIN}
\author{Lea F. Santos}
\affiliation{Department of Physics, Yeshiva University, New York, New York 10016, USA}
\begin{abstract}
This work is concerned with the excited state quantum phase transitions (ESQPTs) defined in Ann.~Phys.~\textbf{323}, 1106 (2008). In many-body models that exhibit such transitions, the ground state quantum phase transition (QPT) occurs in parallel with a singularity in the energy spectrum that propagates to higher energies as the control parameter increases beyond the QPT critical point. The analysis of the spectrum has been a main tool for the detection of these ESQPTs. Studies of the effects of this transition on the system dynamics are more limited. Here, we extend our previous works and show that the evolution of an initial state with energy close to the ESQPT critical point may be extremely slow. This result is surprising, because it may take place in systems with long-range interactions, where the dynamics is usually expected to be very fast. A timely example is the one-dimensional spin-1/2 model with infinite-range Ising interaction studied in experiments with ion traps. Its Hamiltonian has a $U(2)$ algebraic structure. More generally, the slow dynamics described here occurs in two-level bosonic or fermionic models with pairing interactions and a $U(\nu + 1)$ Hamiltonian exhibiting a QPT between its limiting $U(\nu)$ and $SO(\nu + 1)$ dynamical symmetries. In this work, we compare the results for $\nu=1, 2$, and $3$. 
\end{abstract}

\maketitle

\section{Introduction}

Zero temperature quantum phase transitions (QPTs) are driven by quantum fluctuations and take place when a small variation of the Hamiltonian control parameter induces a qualitative change in the ground state properties~\cite{SachdevBook2}. Paradigmatic examples of QPTs include the transitions from a ferromagnet to a paramagnet~\cite{Bitko1996}, from a Mott insulator to a superfluid~\cite{Greiner2002}, and from a normal to a superradiant phase~\cite{Baumann2010}. Studies of QPTs have been at the forefront of theoretical and experimental research in the last few decades, because in addition to the potential to reveal new phases of matter, they may also hold the key to unsolved problems in condensed matter physics, such as high-temperature superconductivity.

Excited state quantum phase transitions (ESQPTs) are generalizations of QPTs  to  excited states. They occur for control parameter values larger than the critical value of the ground state QPT. The ESQPT critical point can then be reached either by varying the control parameter at a constant excitation energy or by increasing the energy at a fixed value of the control parameter. 

The ESQPTs that we address here were introduced in Ref.~\cite{Caprio2008} for many-body models described by dynamical algebras. Signatures of these ESQPTs are traced to specific features of the phase space associated with the system's classical limit. Such signatures include changes in the properties of the wavefunction and in the density of states~\cite{Bernal2008,Caprio2008,Cejnar2008,Cejnar2009,Fernandez2009,Fernandez2011b,Brandes2013,Stransky2014}, which have been experimentally detected in molecules~\cite{Winnewisser2005, Zobov2006,Larese2011, Larese2013}, superconducting microwave billiards~\cite{Dietz2013}, and spinor condensates~\cite{Zhao2014}. Nuclear and molecular systems, as well as integrable and chaotic models have been considered. There has also been interesting attempts to relate the onset of chaos with ESQPTs~\cite{Fernandez2011b}, although this connection may in fact be due to a fortuitous choice of parameters~\cite{Bastarrachea2014b,ChavezARXIV}.

Analyses of how the presence of an ESQPT affects the system dynamics are more restricted~\cite{Fernandez2011,Engelhardt2015}. Yet, they are essential for the potential detection of ESQPTs with experiments where dynamics is routinely studied, such as those with ion traps~\cite{Jurcevic2014,Richerme2014}, Bose Einstein condensates~\cite{Zibold2010}, and nuclear magnetic resonance platforms~\cite{Ferreira2013}. The last two experimental set-ups~\cite{Zibold2010,Ferreira2013}, in particular, reported the observation of the phenomenon of bifurcation, which has been intimately connected with QPTs. As discussed in~\cite{SantosARXIV2}, bifurcation should emerge also due to ESQPTs.

In the present work, we extend our studies~\cite{SantosARXIV2,SantosBernal2015} of the dynamics under a Hamiltonian that exhibits an ESQPT. We have shown that the system evolution can be very slow for initial states with energies very close to the ESQPT critical energy, $E_{\text{ESQPT}}$. Here, we elaborate on this idea by comparing three models and different system sizes.

The system Hamiltonian $\hat {\cal H}$ considered has a $U(\nu+1)$ algebraic structure, with two limiting dynamical symmetries represented by the $U(\nu)$ and the $SO(\nu+1)$ subalgebras~\cite{Cejnar2007},
\begin{equation}
\hat {\cal H} = (1- \xi) \hat {\cal H}_{U(\nu)}  + \dfrac{\xi}{N} \hat {\cal H}_{SO(\nu+1)}  .
\label{ham}
\end{equation}
A second order ground state QPT occurs when the control parameter $\xi$ coincides with $\xi_c = 0.2$ and an ESQPT happens for $\xi > \xi_c$ at  the critical energy $E_{\text{ESQPT}}(\xi)$. 

For $\nu=1$, $\hat {\cal H}$ (\ref{ham}) is the Hamiltonian of the Lipkin-Meshkov-Glick (LMG) model~\cite{Lipkin1965}. It represents the one-dimensional spin-1/2 Hamiltonian with infinite-range interaction and a transverse field~\cite{SantosARXIV2}, which is very close to the system realized with ion traps~\cite{Jurcevic2014,Richerme2014}, where very long-range interactions can be reached. The bosonic form of this Hamiltonian corresponds to the one-dimensional limit of the vibron model, where the bosons represent quanta of molecular vibrations (\emph{vibrons}). The Hamiltonians with $\nu>1$ describe the two-dimensional [$U(3)$] and three-dimensional  [$U(4)$] limits of the molecular vibron model~\cite{Iachello1981,IachelloBook,Iachello1996,Bernal2005,Bernal2008}. In Ref.~\cite{SantosARXIV2}, we studied the effects of an ESQPT on the dynamics of the $U(2)$ Hamiltonian and in Ref.~\cite{SantosBernal2015}, we focused on the $U(4)$ case. Here, we concentrate on the $U(3)$ dynamical algebra to describe the structure of the eigenstates. We show that the states with energy $\sim E_{\text{ESQPT}}$ are highly localized, which affects the system dynamics. We provide a comparison of the system dynamics under $\hat {\cal H} $ with $\nu=1,2$ and $3$.

\section{Model: $U(3)$ Hamiltonian}
The nine generators of the $U(3)$ spectrum generating algebra of the two-dimensional limit of the vibron model  is built from the bilinear products of creation and annihilation operators of two types of bosonic operators: a pair of circular boson operators ($\{\tau^\dagger_\pm, \tau_\pm\}$) and a scalar boson operator  ($\{\sigma^\dagger,\sigma\}$)~\cite{Iachello1996,Bernal2008}. If we assume that the system under study conserves two-dimensional angular momentum, its symmetry algebra is $SO(2)$ and the possible dynamical symmetries are
\begin{eqnarray}
  &U(2)& \, \text{Chain}: \, \, \, \, \, \, U(3)\supset U(2)\supset SO(2) ,\label{chaini}\\
  &SO(3)& \, \text{Chain}: \, \, \, \, \, \, U(3)\supset SO(3)\supset SO(2) .\label{chainii}
\end{eqnarray}

The $U(3)$ Hamiltonian that we consider contains the first order Casimir operator of the $U(2)$ dynamical symmetry, $\hat n$, and the  pairing operator, $\hat P$, as follows~\cite{Bernal2005, Bernal2008}, 
\begin{equation}
\hat {\cal H}_{U(3)} = (1-\xi)\hat n + \frac{\xi}{N-1}\hat P~.
\label{modham1}
\end{equation}
Above, $\hat n = \tau^\dagger_+\tau_++\tau^\dagger_-\tau_-$ is the number of $\tau$ bosons operator, $\hat P=N(N+1)-\hat W^2$, where
\begin{equation}
\hat W^2 =(\hat D_+\hat D_-+\hat D_-\hat D_+)/2+\hat l^2, 
\end{equation}
is the Casimir operator of the $SO(3)$ subalgebra, and $\hat l$ is the angular momentum operator, with
\begin{eqnarray}
&&\hat l = \tau^\dagger_+\tau_+-\tau^\dagger_-\tau_-~, \nonumber \\
&&\hat D_\pm = \sqrt{2}(\pm\tau^\dagger_\pm\sigma-\sigma^\dagger\tau_\mp)~. \nonumber
\end{eqnarray}
The control parameter in Eq.~(\ref{modham1}) takes values between $\xi = 0$ and $\xi = 1$. In the first case, the Hamiltonian is diagonal in the basis of the $U(2)$ Chain, also called the cylindrical oscillator chain, while for $\xi = 1$ the Hamiltonian is diagonal in the $SO(3)$ Chain, which is also known as the displaced oscillator chain~\cite{Bernal2005}.

We  perform our studies in the $U(2)$ basis,
\begin{equation}
\left|\begin{array}{ccccc}
U(3)&\supset& U(2)&\supset& SO(2)\\
\left[N\right]   &       & n   &       & \ell
\end{array}\right\rangle ~~,
\label{cobas}
\end{equation}
where  the quantum number $N$ is the system size, that is the total number of $\sigma$ and $\tau$ bosons. $N$ labels the totally symmetric representation of $U(3)$ that spans the system's Hilbert space. The branching rules in the $U(2)$ basis are
\begin{align}
n & = N, N-1, N-2, \ldots, 0 , \nonumber
\end{align} 
and
\begin{align}
\ell & = \pm n, \pm (n-2), \ldots, \pm 1 \,[\mbox{or} \,\, 0]~,~~ \mbox{if} ~~ n = \mbox{odd} \,\,[\mbox{or even} ]\nonumber
\end{align} 

The $U(2)$ basis has a simple expression in terms of the $\sigma$ and $\tau$ creation boson operators, 
\begin{equation}
\left|[N]n \ell \right\rangle  =
{\cal N}^N_{n\ell}\left(\sigma^\dagger\right)^{N-n} \left(\tau^\dagger_{+}\right)^\frac{n+\ell}{2}
\left(\tau^\dagger_{-}\right)^\frac{n-\ell}{2}\left|0\right\rangle~~,
\label{u3state}
\end{equation}
\noindent where the normalization constant is
\begin{equation}
{\cal N}^N_{n\ell}=\frac{1}
{\sqrt{(N-n)!\left(\frac{n+\ell}{2}\right)!\left(\frac{n-\ell}{2}\right)!}} ~~.
\end{equation}
The Hamiltonian matrix elements in this basis are given by~\cite{Iachello1996,Bernal2008}
\begin{eqnarray}
&&\langle [N] n_2 \ell |\hat n| [N] n_1 \ell\rangle = n_1\, \delta_{n_2,n_1} \label{nmatel_u2}\\
&&\langle [N] n_2 \ell |{\hat P}| [N] n_1\ell\rangle = \\
&&N(N+1) \!-\! \left[(N \!-\! n_1)(n_1 \!+\! 2) \!+\! (N \!-\! n_1 \!+\! 1)n_1 \!+\!  \ell^2\right] \delta_{n_2,n_1} \nonumber\\
&&+ \sqrt{(N-n_1+2)(N-n_1+1)(n_1+\ell)(n_1-\ell)}\,\delta_{n_2,n_1-2}  \nonumber\\
&&+
\sqrt{(N \!-\!  n_1)(N \!-\!  n_1 \!-\!  1)(n_1 \!+\!  \ell+2)(n_1 \!-\!  \ell \!+\!  2)}\,\delta_{n_2,n_1 + 2}~~.\nonumber
\end{eqnarray}
The conservation of angular momentum makes the resulting matrix
block-diagonal, so the blocks corresponding to different $\ell$ values
can be diagonalized separately.

\section{Localization at the ESQPT critical point}

\begin{figure}[h]
\includegraphics[width=\columnwidth]{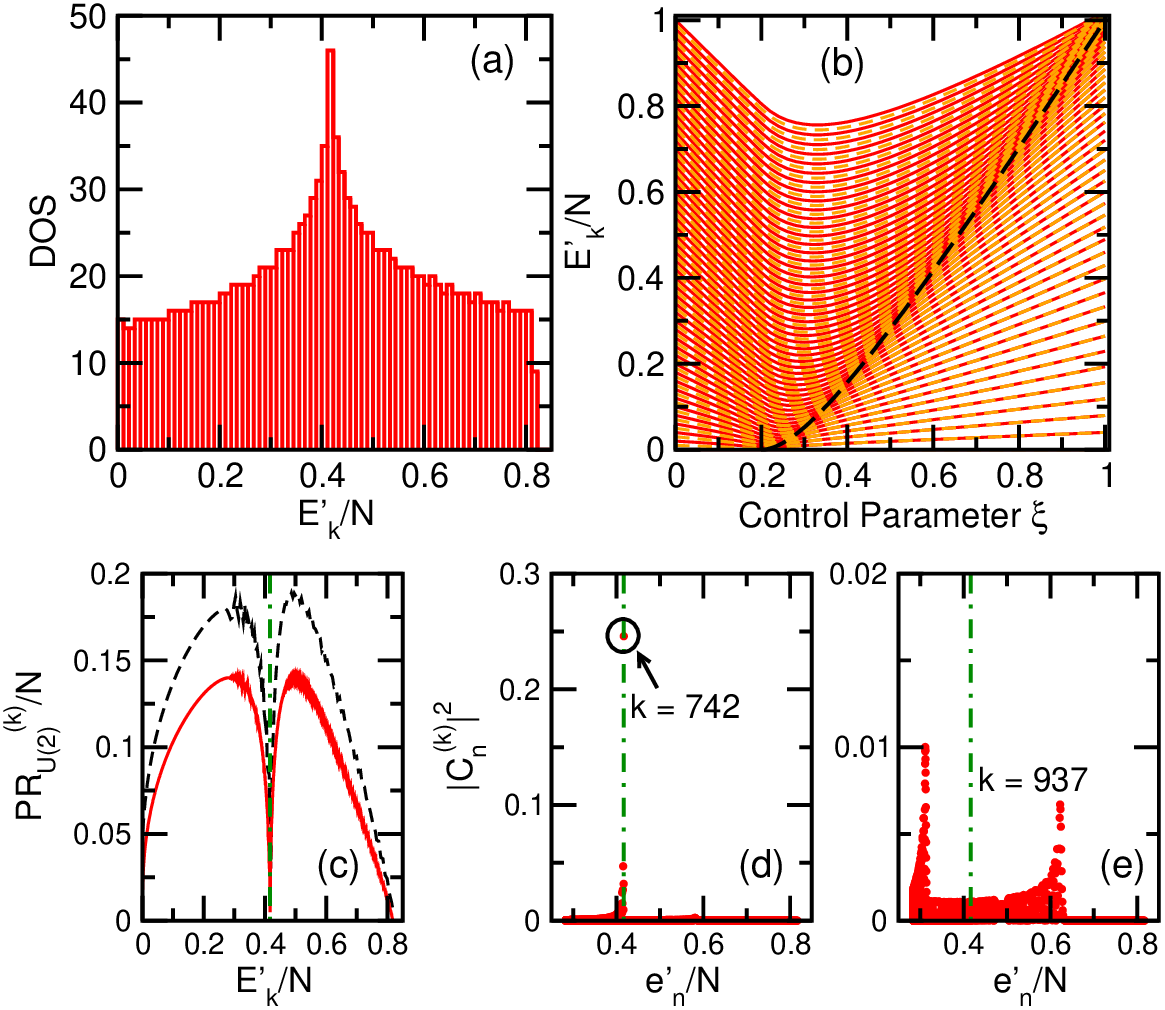}
  \caption{\label{Fig01}
Panel (a): Normalized density of states for $\xi=0.6$, $N=3000$, $\ell=0$. 
Panel (b): Normalized excitation energies \textit{vs} $\xi$ for the $U(3)$ total Hamiltonian [Eq.~(\ref{modham1})]; $N = 100$, $\ell=0$ (full red lines) and $\ell=1$ (dashed orange lines).  The separatrix [Eq.~(\ref{eq:separatrix})] is indicated with the dashed line.
Panel (c): Participation ratio of all eigenstates written in the $U(2)$ basis; $\xi=0.6$, $\ell = 0$, and $N=300$ (dark curve) and $N=3000$ (light curve). The vertical line marks $E_{\text{ESQPT}}$ obtained from Eq.~(\ref{eq:separatrix}).
Panels (d) and (e): squared coefficients $|C_{n}^{(k)}|^2$ of the eigenstates $|\psi_k\rangle$ written in the $U(2)$ basis  {\em vs} the energies of the corresponding basis vectors. The eigenstate in (d) is the closest one to the ESQPT critical point and it has $E_k'/N = 0.4168$, and the eigenstate in (e) has energy $E_k'/N = 0.4872$; $\xi =0.6$, $N = 3000$, $\ell=0$. Arbitrary units. }
\end{figure}

A commonly employed way to detect the presence of an ESQPT is via the computation of the density of states, which diverges at the critical energy $E_{\text{ESQPT}}$, as seen in Fig.~\ref{Fig01} (a). We notice that in our figures, the eigenvalues $E_k$ are shifted by the energy of the ground state $E_0$, that is $E_k'= E_k-E_0$, and they are normalized by $N$.

The clustering in Fig.~\ref{Fig01} (a) is equivalently captured by Fig.~\ref{Fig01} (b), which shows all the eigenvalues as a function of the control parameter, from $\xi=0$ to $\xi=1$. For $\xi=\xi_c$, there is a high concentration of eigenvalues at $E_k'/N=0$, which gets displaced to higher values, $E_k'/N=E_{\text{ESQPT}}$, as $\xi$ further increases. This is indicated with the separatrix (dashed line), that marks the ESQPT. The equation for the separatrix can be obtained in the semiclassical limit~\cite{Bernal2008,Bernal2010} and is given by 
\begin{equation}
E_{\text{ESQPT}}  (\xi)= \frac{\left(1-5\,\xi_{\text{ESQPT}}\right)^2}{16\,\xi_{\text{ESQPT}}}.
\label{eq:separatrix}
\end{equation}

\subsection{Eigenstates}

Figure~\ref{Fig01} (b) displays eigenvalues with two angular momenta, $\ell=0$ and $\ell=1$. It is noticeable that states having different angular momentum with energies below the separatrix are degenerate, while this degeneracy is broken for energies above $E_{\text{ESQPT}}$. Thus, the separatrix marks a change in the structure of the eigenstates. Those states with $E_k'/N < E_{\text{ESQPT}}$ are closer to the eigenstates of the $SO(3)$ Chain, while above the separatrix, they approach the eigenstates of the $U(2)$ Chain \cite{Caprio2008}.  We verified that this change is reflected in the level of delocalization of the eigenstates $|\psi_{k}\rangle$ written in the $U(2)$ basis, 
\begin{equation}
|\psi_{k}\rangle = \sum_{n=\ell}^{N} C_n^{(k)} |[N] n \ell\rangle_k,
\end{equation} 
where the sum involves either even or odd values of $n$, depending on the $\ell$ parity.

The spreading of an eigenstate in a chosen basis is quantified, for instance, with the participation ratio (PR) \cite{Santos2005loc,Gubin2012,Torres2013}, which is defined as
\begin{equation}
\text{PR}^{(k)}_{U(2)} =\frac{1}{\sum_{n} | C_{n}^{(k)}|^4}~.
\label{eq:IPR}
\end{equation}
An extended state has a large value of PR, while for $\xi=0$, $\text{PR}^{(k)}_{U(2)} =1$ for every $k$, since the eigenstates coincide with the $U(2)$ basis vectors.

Figure~\ref{Fig01} (c) shows the values of the PR for all $\ell = 0$ eigenstates for $\xi=0.6$ and two different system sizes. In addition to being small at the borders of the spectrum, the PR also dips abruptly at $E_{\text{ESQPT}}$. At first sight this might be surprising, since this is where the divergence of the density of states occurs [cf. Fig.~\ref{Fig01} (a)], so strong mixing and thus delocalized states could have been expected. However, the separatrix marks the point where the eigenstates  transition from one dynamical symmetry to the other. Starting at $E_k'/N=0$ and going up in energy, the point where $E_k'/N = E_{\text{ESQPT}}$ marks the transition from the $SO(3)$ symmetry to the $U(2)$ symmetry. At this point, the eigenstates become highly localized in the ground state of the $U(2)$ part of the Hamiltonian, that is in the state $|[N] 0 0\rangle$. This strong localization can also be understood from classical considerations, as further discussed in Ref.~\cite{SantosARXIV2}. This sudden change in the value of the PR serves as an alternative method to detect an ESQPT.

To confirm that the localization occurs in the ground state of the $U(2)$ part of $\hat {\cal H}$, we show in Fig.~\ref{Fig01} (d), the components $| C_n^{(k)}|^2$ as a function of the normalized energies of the $U(2)$ basis vector, 
\begin{equation}
e'_n/N = (\langle [N] n \ell | \hat{\cal H} | [N] n \ell\rangle - E_0)/N ,
\end{equation} 
for the eigenstate that is closest to the separatrix. It is indeed highly localized at  $e'_0/N \sim E_{\text{ESQPT}}$. To contrast this result, we show in Fig.~\ref{Fig01} (e), the structure of an eigenstate with $E_k'/N > E_{\text{ESQPT}}$. The state is delocalized and shows no preference on any $U(2)$ basis vector.

\subsection{Fidelity and number of $\tau$ bosons}

The ground-state fidelity, $F$, is often used to detect a QPT~\cite{Zanardi2006}. It corresponds to the overlap between two ground states obtained for two values of the control parameter that differ by a very small amount $\delta$,
\begin{equation}
F = \langle \psi_0 (\xi) |\psi_0 (\xi + \delta) \rangle .
\end{equation}
Away from the QPT critical point, the two ground states are very similar and $F \sim 1$, but at $\xi_c$, the ground-state fidelity shows a sudden drop. 

\begin{figure}[h]
  \includegraphics[width=\columnwidth]{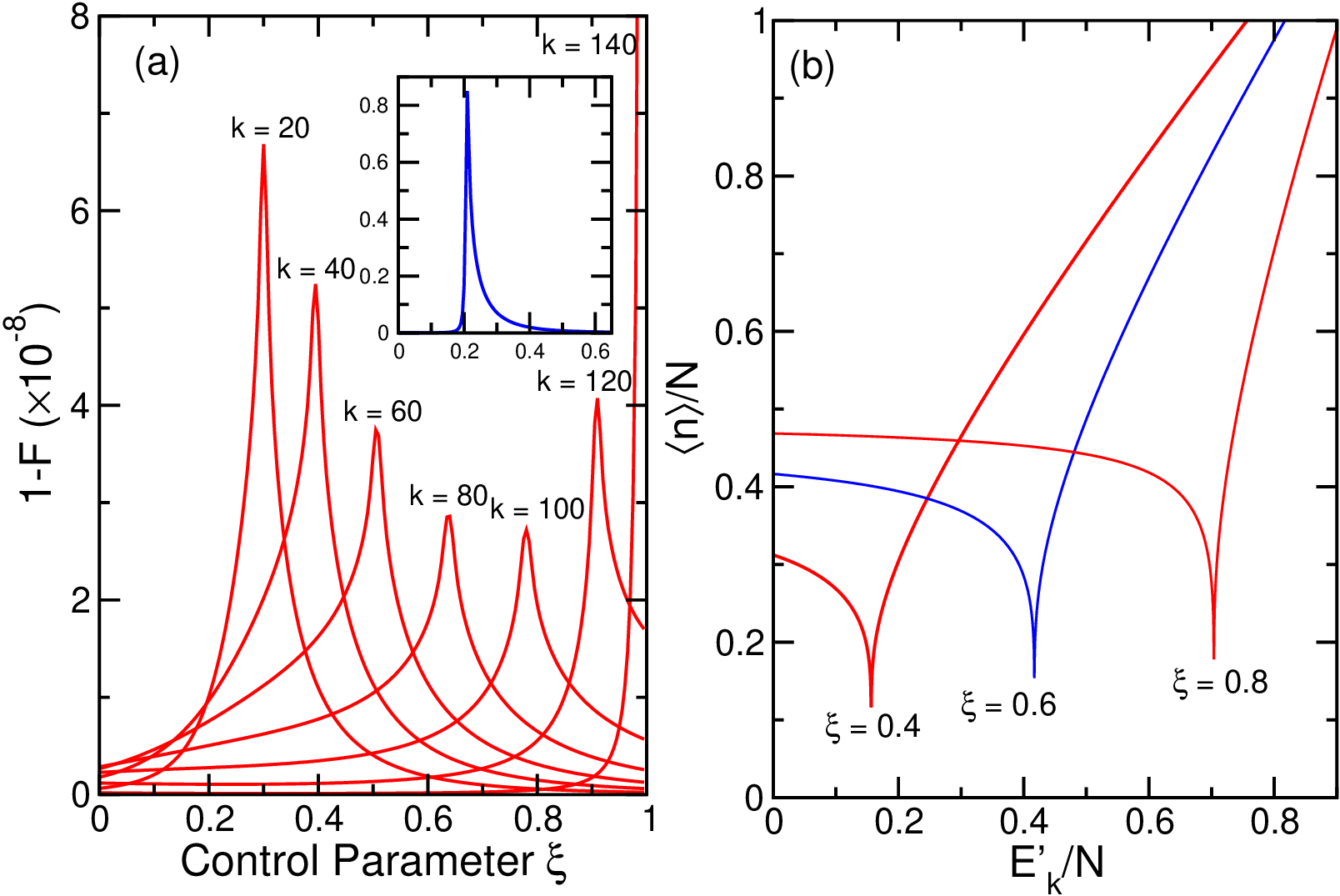}
  \caption{\label{Fig02}
Panel (a): $1-F$ \textit{vs} $\xi$ for the $U(3)$ model Hamiltonian [Eq.~(\ref{modham1})]; $N=300$, $\ell=0$. The values of $k$ for the chosen excited eigenstates are indicated in the figure. For comparison, the inset shows  $1-F$  \textit{vs} $\xi$ for the ground state.
Panel (b): Expectation value of the number operator for all eigenstates with $\xi = 0.4, 0.6$, and $0.8$; $N=3000$, $\ell=0$. Arbitrary units.}
\end{figure}

The same idea carries over to ESQPTs, except that the abrupt change in $F$ now occurs for two excited states with energies $\sim E_{\text{ESQPT}}$. It reflects the change in the structure of the eigenstates across the separatrix. This is illustrated in Fig.~\ref{Fig02} (a), where we show $1-F$ {\em vs} $\xi$ for some chosen eigenstates. For each one, $1-F$ peaks at the value of $\xi$ for which  $E_k'/N \sim E_{\text{ESQPT}}$. 

The strong localization of the eigenstates at the separatrix in the basis vector with $n=0$ causes the expectation value of the $\tau$ boson number operator, $\langle \psi_k |\hat{n} | \psi_k \rangle$, to have a pronounced minimum in the critical energy~\cite{Bernal2010}, as seen in Fig.~\ref{Fig02} (b). Thus, in addition to PR, both $\langle \psi_k |\hat{n} | \psi_k \rangle$ and the fidelity are alternative quantities capable of detecting the appearance of an ESQPT. 

\section{Dynamics}

Since the eigenstates close to the separatrix are very localized in $|[N] 00\rangle$, the evolution of this basis vector under the Hamiltonian (\ref{modham1}) must be much slower than the dynamics of other basis vectors, as we  show in this section. The speed of the evolution of $U(2)$ basis vectors is therefore another way to detect the presence of an ESQPT.

To examine how fast an initial state $|\Psi(0)\rangle= | [N] n \ell\rangle$ changes in time, we consider the survival probability~\cite{Borgonovi2016},
\begin{eqnarray}
S_p(t) &\equiv& \left| \langle \Psi(0) | e^{-i \hat{\cal H} t} | \Psi(0) \rangle \right|^2  = \left| \sum_k |C_{n}^{(k)} |^2 e^{-i E_k t}  \right|^2 \nonumber \\
&=&\left| \int  dE e^{-i E t} \rho_{n}(E) \right|^2,
\label{eq:fidelity}
\end{eqnarray} 
where $\rho_{n}(E) = \sum_k |C_{n }^{(k)} |^2\delta(E-E_k)$ is the energy distribution of the initial state $|[N] n \ell\rangle$ weighted by the components   $|C_{n}^{(k)} |^2$.  This distribution is referred to as local density of states (LDOS) \cite{Borgonovi2016}. Notice that the survival probability is the absolute square of the Fourier transform of the LDOS.

\begin{figure}[h]
  \includegraphics[width=\columnwidth]{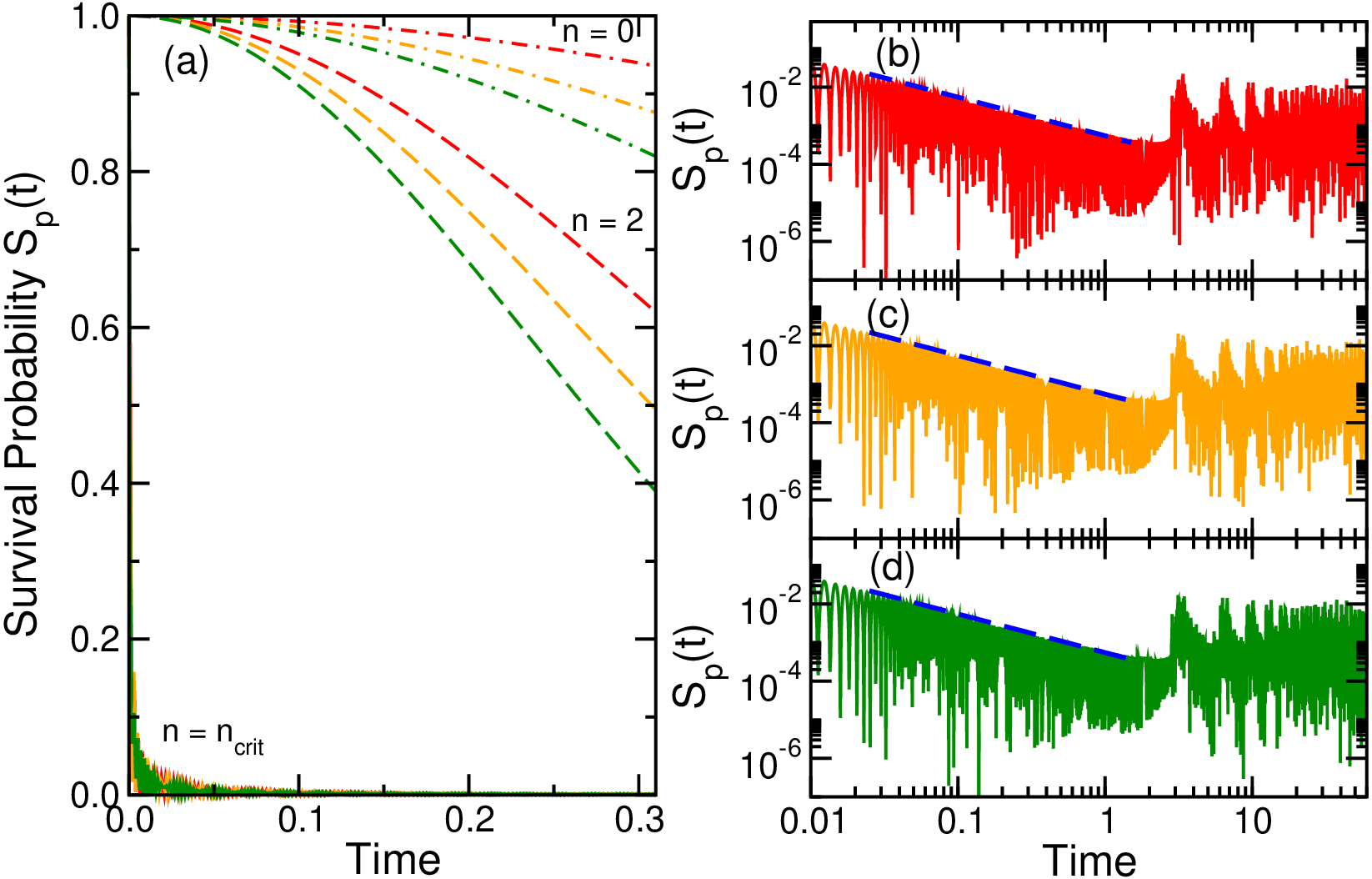}
  \caption{\label{Fig03}
Survival Probability {\em vs} time; $\xi=0.6$, $N=5000$. In (a), the initial states from top to bottom have: $n=0$ for the $U(2)$, $U(3)$, and $U(4)$ Hamiltonians;  $n=2$ for the $U(2)$, $U(3)$, and $U(4)$  Hamiltonians; and $n=n_c$ (the closest $e'_{n}/N$ to $E_{\text{ESQPT}}$, apart from $e'_{0}/N$) for the $U(2)$, $U(3)$, and $U(4)$ Hamiltonians. Panels (b), (c), and (d) show the long-time evolution of $|\Psi(0)\rangle= |[N] n_c 0\rangle$  for $U(2)$, $U(3)$, and $U(4)$, respectively; dashed lines give $F(t) \propto 1/t$. Arbitrary units.}
\end{figure}

In Fig.~\ref{Fig03} (a), we compare the evolution of $|[N] 00\rangle$, $|[N] 2,0\rangle$, and $|[N] n_c 0\rangle$ under the $U(\nu)$ Hamiltonian (\ref{ham}) with $\nu=1,2$ and $3$. The initial state  $|\Psi(0)\rangle= |[N] n_c 0\rangle$ is the state with the closest value of $e'_{n_c}/N$ to $E_{\text{ESQPT}}$, with $n_c\ne 0$. State $|[N] 00\rangle$ has $e'_0/N\sim E_{\text{ESQPT}}$, because it is highly localized in the eigenstate at the separatrix. As $n$ increases, the states $|[N] n 0\rangle$ get more and more delocalized in the energy eigenbasis. The LDOS for $|[N] n_c 0\rangle$ is approximately symmetrically distributed between eigenstates with $E_k'/N < E_{\text{ESQPT}}$ and eigenstates with $E_k'/N > E_{\text{ESQPT}}$, which guarantees that $e'_{n_c}/N$ is closer to $E_{\text{ESQPT}}$ than $e'_{2}/N$ \cite{SantosBernal2015,SantosARXIV2}.

Figure~\ref{Fig03} (a) shows that the evolution of $|[N] 00\rangle$ is significantly slower than that for the other basis vectors. Overall, the dynamics also accelerates as $\nu$ increases, but in a similar rate for all initial states, which assures the distinctly slow behavior of $|[N] 00\rangle$ when compared to other basis vectors.

Figures~\ref{Fig03} (b), (c), and (d) depict the long-time decay of the survival probability for the delocalized state $|[N] n_c0\rangle$ for $\nu=1,2$ and $3$, respectively. A powerlaw behavior $\propto 1/t$ is found for all three cases. Interestingly, this is also the behavior observed for the integrable XX model with a single excitation~\cite{SantosARXIV2}, which is a spin-1/2 noninteracting model with only nearest-neighbor couplings. The analysis of the LDOS for $|[N] n_c0\rangle$ reveals a shape $\propto (\text{constant} - E^2)^{-1/2}$, whose Fourier transform indeed leads to $S_p(t) \propto 1/t$. This shape is also found for the LDOS and for the density of states of the XX model with a single excitation. This suggests a relationship between the short-range XX model and the infinite-range LMG model, which deserves further investigation. Similarities between these two models with respect to the ground state entanglement entropy have been found in~\cite{Latorre2005,Barthel2006}.

\begin{figure}[h]
  \includegraphics[width=\columnwidth]{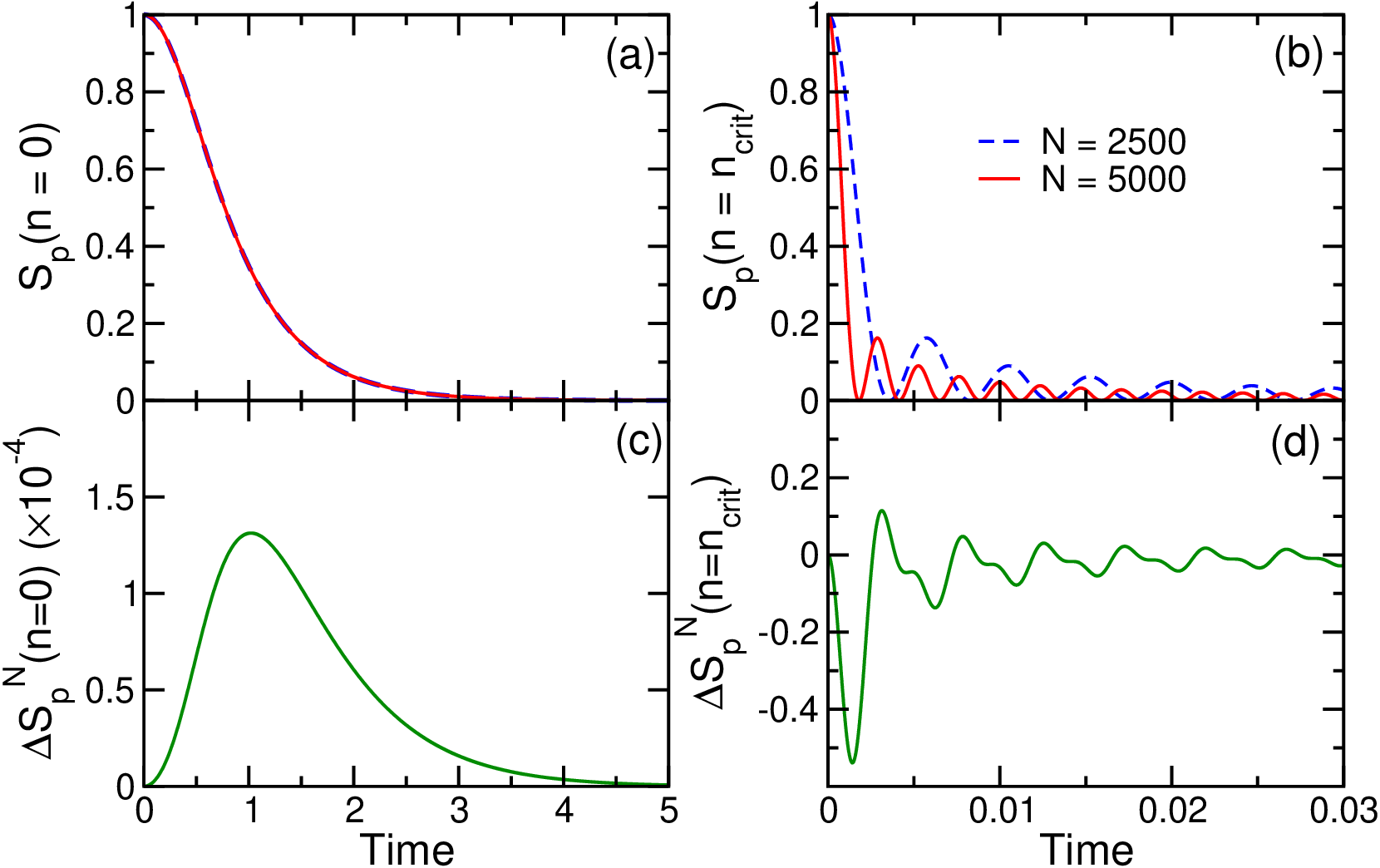}
  \caption{\label{Fig04}
Top panels (a) and (b): Survival probablity {\em vs} time for two different system sizes; $N=2500, 5000$. Bottom panels (c) and (d): Difference between the survival probability for the same two system sizes \textit{vs} time. In (a) and (c): $\langle n \rangle=0$; in (b) and (d): $\langle n \rangle=n_c$ (the second closest $e'_{n}/N$ to $E_{\text{ESQPT}}$). The total Hamiltonian has a $U(3)$ algebraic structure [Eq.~(\ref{modham1})] with $\xi = 0.6$ and $\ell = 0$. Arbitrary units. }
\end{figure}

A comparison of the decay of the survival probability for different system sizes is provided in Fig.~\ref{Fig04} for $|[N] 00\rangle$ (a,c) and $|[N] n_c0\rangle$ (b,d). In Fig.~\ref{Fig04} (a), one sees that before the appearance of oscillations caused by the finite system size,  $S_p(t)$ practically coincides for different values of $N$. A closer look, in Fig.~\ref{Fig04} (c), at the difference $\Delta S_p^N (t) = S_p^{N_2} (t) - S_p^{N_1} (t)$ for two different system sizes $N_1$ and $N_2$, where $N_2>N_1$, reveals that the decay is slightly slower for larger $N$. In contrast, the evolution of $|[N] n_c0\rangle$ clearly accelerates as $N$ increases, which is evident already in Fig.~\ref{Fig04} (b) and further reinforced by the negative values of the difference $\Delta S_p^N (t)$ in Fig.~\ref{Fig04} (d). The slow behavior of  $|[N] 00\rangle$ is therefore a real effect of the presence of the ESQPT and not a finite size effect. 

\section{Conclusions}

We have shown that the presence of an ESQPT can be detected by comparing the speed of the evolution of different $U(\nu)$ basis vectors propagating under a $U(\nu+1)$ Hamiltonian. In the particular case of $\nu=1$, the $U(2)$ Hamiltonian coincides with the spin-1/2 model with infinite-range Ising interaction in the $z$-direction and a transverse field in the $x$-direction. For this model, the initial state that shows a very slow dynamics at the ESQPT critical point is the one where all spins point down in the $z$-direction, which is a state routinely considered in the studies of quench dynamics performed with trapped ions~\cite{Jurcevic2014,Richerme2014}.  

Our findings demonstrate that the fast evolution expected to be seen in systems with long-range interaction is not a general rule; slow dynamics may be found as well. The behavior of a system out of equilibrium depends on the interplay between its Hamiltonian and its initial state. This has been discussed in Ref.~\cite{SantosARXIV} in the context of systems with long-range interaction and has been a recurrent message of our works about nonequilibrium quantum dynamics~\cite{Torres2014PRA,Torres2014PRE,Torres2014PRAb,Torres2014NJP,TavoraARXIV} and thermalization~\cite{Torres2013}.


\section{Acknowledgments}
FPB was funded by MINECO grant FIS2014-53448-C2-2-P and by Spanish Consolider-Ingenio 2010 (CPANCSD2007-00042). LFS was supported by the NSF grant No.~DMR-1147430.  We thank Alejandro Frank for his hospitality at the Centro de Ciencias de la Complejidad (C$_3$) of the UNAM in Mexico, where part of this work was carried out.


\end{document}